\documentclass[a4paper,fleqn,usenatbib]{mnras}

\usepackage{mathptmx}
\usepackage[T1]{fontenc}
\usepackage{ae,aecompl}

\usepackage{graphicx}	% Including figure files
\usepackage{amsmath}	% Advanced maths commands
\usepackage{amssymb}	% Extra maths symbols

\newcommand{\dnu}{$\Delta\nu$}
\newcommand{\numax}{$\nu_{\rm max}$}
\newcommand{\nuac}{$\nu_{\rm ac}$}

\title[A new reference function for the \dnu\ scaling relation]{Significantly improving stellar mass and radius estimates: A new reference function for the \dnu\ scaling relation}

\author[E. Guggenberger et al.]{
Elisabeth Guggenberger,$^{1,2}$\thanks{E-mail: guggenberger@mps.mpg.de}
Saskia Hekker,$^{1, 2}$
Sarbani Basu,$^{3}$
Earl Bellinger $^{1, 2}$
\\
$^{1}$Max Planck Institut f\"ur Sonnensystemforschung, Justus-von-Liebig-Weg 3, 37077 G\"ottingen, Germany\\
$^{2}$Stellar Astrophysics Centre, Dept. of Physics and Astronomy, Aarhus University, Ny Munkegade 120, 8000 Aarhus C, Denmark\\
$^{3}$Department of Astronomy, Yale University, 52 Hillhouse Avenue, New Haven, CT 06511, USA
}

\date{Accepted XXX. Received YYY; in original form ZZZ}

\pubyear{2016}

\begin{document}
\label{firstpage}
\pagerange{\pageref{firstpage}--\pageref{lastpage}}
\maketitle

\begin{abstract}

The scaling relations between global asteroseismic observables and stellar properties are widely used to estimate masses and radii of stars exhibiting solar-like oscillations. Since the mass and radius of the Sun are known independently, the Sun is commonly used as a reference to scale to. However, the validity of the scaling relations depends on the homology between the star under study and the reference star. Solar-like oscillators span a wide range of masses and metallicities, as well as evolutionary phases. Most of these stars are therefore not homologous to the Sun. This leads to errors of up to 10\% (5\%) in mass (radius) when using the asteroseismic scaling relations with the Sun as the reference. In this paper we derive a reference function to replace the solar-reference value used in the large-frequency-separation scaling relation. Our function is the first that depends on both effective temperature and metallicity, and is applicable from the end of the main sequence to just above the bump on the red giant branch. This reference function improves the estimates of masses and radii determined through scaling relations by a factor of 2, i.e. allows masses and radii to be recovered with an accuracy of 5\% and 2\%, respectively.
\end{abstract}


%%%%%%%%%%%%%%%%%%%%%%%%%%%%%%%%%%%%%%%%%%%%%%%%%%

%%%%%%%%%%%%%%%%% BODY OF PAPER %%%%%%%%%%%%%%%%%%

\section{Introduction}

Mass and radius are the most fundamental among stellar parameters. Knowledge of
these parameters not only enables one to characterize a star, but it is also a
prerequisite for characterizing planets orbiting a star.
%The most reliable method to derive a given star's mass and radius from its asterosesismic signal is the extraction of individual frequencies (peak-bagging) and dedicated modeling (boutique modeling). But this method is computationally expensive and time consuming and hence feasible only for small numbers of stars. 
%maybe this can come later, in the problems section...
Asteroseismic scaling relations are a fast and straightforward method that
can be used to estimate masses and radii of stars that exhibit solar-like
oscillations. These are low-amplitude oscillations that are excited
stochastically by convection in the outer zones of cool stars.

One scaling 
relation links the so-called large frequency separation, i.e.,
the frequency difference between modes of the same degree and consecutive
radial orders (\dnu) to the square root of the stellar mean density
\citep{ulrich86}.

\begin{equation}
{\Delta\nu}\propto \sqrt{\frac{M}{R^3}},
\label{dnu}
\end{equation}
where $R$ and $M$ are the stellar radius and mass. The
constant of proportionality is determined using \dnu\ for the Sun,  $\Delta\nu_{\odot}=135.1 \pm 0.1\mu $Hz
\citep{hub11}.

The second scaling relation links the frequency at  which the maximum of the
oscillation power occurs, \numax, with stellar properties. Based on the argument by \citet{bgnr91} that
\numax\ should scale with the acoustic cutoff frequency \nuac, \citet{kb95}
formulated the \numax\ scaling relation.
\begin{equation} 
\nu_{\rm{max}}\propto \frac{M}{R^2\sqrt{T_{\rm{eff}}}}.
\label{numax}
\end{equation}
Again, the constant of proportionality is determined using the solar 
value of \numax\ as reference:
 $\nu_{\rm{max}, \odot}  =  3050$  \citep{kb95}, and  $T_{\rm{eff}, \odot}=5772$ K \citep{mam15}.

The accuracy of the stellar
radii obtained from scaling relations has been tested through comparisons with
radii obtained from independent measurements such as interferometry. These
estimates are in good agreement within uncertainties, which can be about 10\% for
interferometric radii of red giants due to uncertainties in the parallaxes
\citep{hub12}.  
However, the differences in structure between the Sun and the stars 
investigated result in inherent systematic deviations of the values computed with the scaling relations with respect to the true values.
The underlying physical reason of the \dnu\ relation is understood. Hence, we can compute the deviation of \dnu\ from the scaling relation with respect to the \dnu\ value obtained from individual frequencies (see Fig.~\ref{fig:fit}). The influence of \numax\ on the \dnu\ scaling relation can currently not be studied through comparisons with models because \numax\ can currently not be calculated theoretically.

Using the \dnu\ scaling relation with a fixed reference value results in masses and radii with an intrinsic error
of about 10 and 5 percent, respectively. 
\subsection{Mismatches and earlier corrections}
In their study of a planet-hosting red-giant star \citet{hub13} noticed that
results from grid-based modeling using the scaling relations indicated in
Eqs~\ref{dnu} and \ref{numax} and the solar reference differ by 5\% from those
based on the modeling of individual frequencies.  Mismatches for non-solar
metallicities are also known: in their sample of metal-poor stars (-2.3 dex <
[M/H] < -1.0 dex) \citet{ep14} found masses derived from scaling relations to
be systematically higher than expected from other astrophysical priors.

\begin{figure}
	\includegraphics[width=\columnwidth]{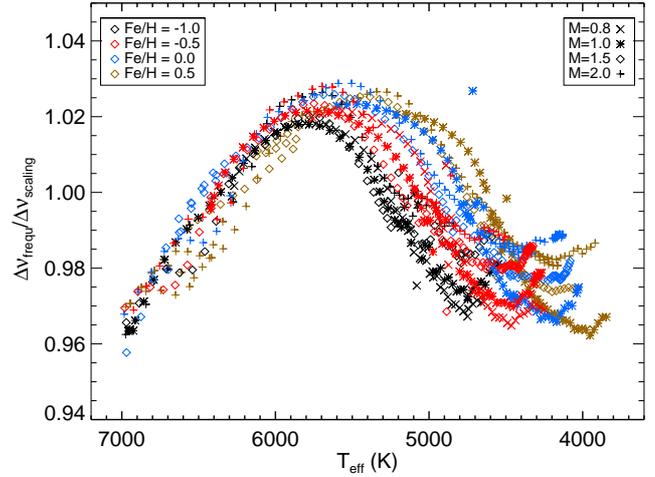}
    \caption{Ratio of $\Delta\nu$ derived from individual frequencies to
$\Delta\nu$ determined using the $\Delta\nu$ scaling relation (Eq.~\ref{dnu}) with a solar reference value.
Different metallicities are shown in different colors, and different masses
with different symbols (see legends). The horizontal dashed line indicates a
ratio of 1.} \label{fig:fit}
\end{figure}

Efforts have been made by several groups to improve the accuracy of the
scaling relations. \citet{belk13} and \citet{moss13} argued that observations
are often only possible far from the asymptotic regime, i.e., at too low radial
orders to justify the asymptotic approximation. Therefore they suggested to use the second
order effects in the large frequency separation to derive asymptotic spacings
from the observations. It was then shown by \citet{hekk13} that the postulated
discrepancies from the asymptotic behaviour have most likely been
overestimated.  \citet{mig12} suggested a correction factor of 2.7 percent for
the radius of stars in the red clump. This correction factor was derived based
on the fact that two models with the same mass and radius show a significant
difference in \dnu\ depending on their evolutionary state. This difference in
\dnu\ between red-clump stars and red giant branch stars is due to differences
in the sound-speed profile. A different approach was taken by \citet{white11}
who proposed a quadratic correction function $f(T_{\rm{eff}})$ that was
calibrated with \dnu\ values derived from the pulsation frequencies of 
stellar models, in particular with models constructed with the ASTEC code \citep{jcd2008}.
Their correction is applicable from the main-sequence to temperatures down to 4700K which includes sub giants and red giants up to near the RGB bump depending on mass and metallicity.
 As can be seen from Fig.~\ref{fig:fit}, the deviation from the
scaling relation cannot be represented as a quadratic function for
 temperatures below this limit, which correspond to temperatures of many of the observed red giants. Additionally, the correction was only calibrated for stars with near-solar metallicities, and no [Fe/H] dependence is included in their formulation.
Given the astrophysical importance of cool red-giant stars and the fact that
many stars have metallicities outside the range of the models on which the previous correction was based (the metallicities used were Z=0.011, Z=0.017 and Z=0.028, corresponding to [Fe/H] values of -0.19, 0.0, and 0.21), 
 our goal is to determine a reference function that works for red giants, and that
can be applied to stars in a wider range of metallicities. 
Recently another approach has been suggested by \citet{sharma16} which relies on interpolation in a grid of models. This method however requires the use of grids of models, while we present here a method that can be straightforwardly applied.

\begin{figure}
	\includegraphics[width=\columnwidth]{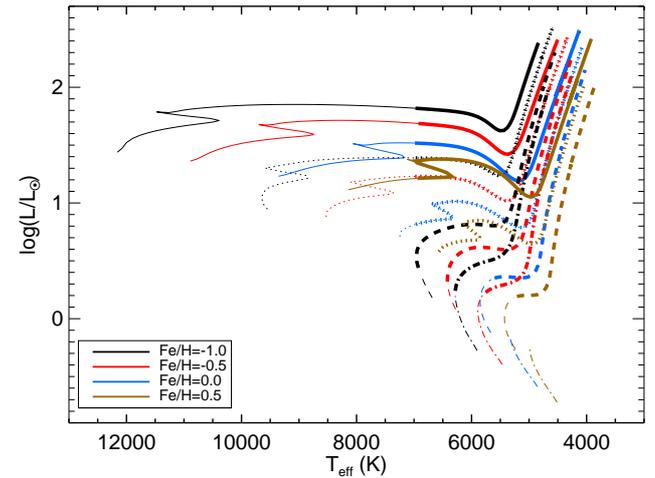}
    \caption{Hertzsprung-Russell diagram of the YREC models used in this study.
Different metallicities are shown in different colors (see legend), while
masses are indicated in different line styles (solid: M=2 M$_\odot$, dotted:
M=1.5 M$_\odot$, dashed: M=1 M$_\odot$, dash-dotted M=0.8 M$_\odot$). The models
actually used in this study (see Section 3.1) are indicated with thick lines.}
    \label{fig:HRD}
\end{figure}

\section{Models}
We used models computed with the YREC stellar evolution code \citep{dem08}.
Models were computed using OPAL opacities \citep{ir96} supplemented with low
temperature (log T < 4.1) opacities of \citet{ferg05} and the OPAL equation of
state \citep{rog02}. All nuclear reaction rates are obtained from
\citet{adel98}, except for that of the $^{14}N(p,\gamma)^{15}O$ reaction, for
which we use the rate of \citet{for04}. We computed models with masses of 0.8,
1.0, 1.5, and 2.0 M$_\odot$, and metallicities of $-1.0$, $-0.5$, $0.0$, and
$0.5$ dex. Although this is a coarser grid than used by \citet{sharma16} we deem this sufficient based on the monotonic behavior of the models with metallicity. Evolution was followed from the main sequence to slightly beyond the bump on the
red-giant branch. We included models with \numax\ > 6 $\mu$ Hz. See Fig.~\ref{fig:HRD} for a Hertzsprung-Russell diagram showing the model tracks.
All models described here were constructed with the Eddington $T-\tau$
relation in their atmosphere. To study the effect of atmospheric structure, we
use two sequences of solar metallicity and masses of $M$=1.0; 1.2~M$_{\odot}$
models constructed with the Krishna Swamy $T-\tau$ relation
\citep{krishnaswamy1966}, where the $M$=1.0~M$_{\odot}$ sequence also has a
slightly different mixing length parameter.

\subsection{Calculating  $\Delta\nu$}
We derive \dnu\ from the computed oscillation radial-mode frequencies of the models.
Since frequencies of any order can be computed, including those orders that
are not observed, we use only a subset of the modes.
We selected modes around the value of  \numax.
The value of \numax\ was computed for each model from the known mass,
temperature and radius using Eq.~\ref{numax} with $\nu_{\rm max, \odot} =
3050~\mu$Hz  \citep{kb95};  \dnu\ was then determined by a weighted linear fit
to the frequencies versus the radial order. This is similar to the method of \citet{white11}. However, we used a different width of the weighting function. The weights were set using a
Gaussian centered at \numax\ with a height of 1 and a full-width at half maximum
 set to $5\Delta\nu$. The width is defined empirically and taken such that it is applicable over a wide parameter space to include enough frequencies to determine $\Delta\nu$ and to resemble the number of radial modes typically observed. For red giants this width is consistent with the observed FWHM of $(0.66\pm0.01)\nu_{\rm max}^{0.88 \pm 0.01}$ \citep{mosser2012}. Our definition of the width of the weighting function leads to a wider gaussian on the red giant branch compared to the definition FWHM=$0.25 \cdot \nu_{max}$  which was used by \citet{white11}. 
The $\Delta\nu$ value used to determine the width of the Gaussian function is
computed using Eq.~\ref{dnu} with $\Delta\nu_{\odot} =135~\mu$Hz. It was shown by \citet{hekk13} that the \dnu\ values of \textit{Kepler} red giants derived from the radial modes do not differ significantly from the values obtained from the power spectrum of the power spectrum (see their Fig. 1). It is therefore safe to assume that our method provides a good estimate of \dnu\ even when mixed modes are present in the power spectrum. 

As a Gaussian
weighting function assigns non-zero weights even at frequencies far away from
its centre, it could  include signal from above the acoustic cutoff frequency
\nuac\  if oscillations are computed in that regime. To mitigate this, the dimensionless value of the normalized weighting function at \nuac\ was subtracted from the 
weighting function so that it was shifted downwards to become zero at \nuac.
Shifts were small ($10^{-9}$ to $10^{-5}$) on the main sequence and larger (up
to 0.5) on the red-giant branch. Negative values of the shifted weighting
function were set to zero.  

We note that we do not have a calibrated solar model in our set. 
Our most solar like model has a \dnu\ value of 136.1~$\mu$Hz. The offset between the observed solar value and this value is most likely due to surface effects that cannot be incorporated directly in the current analysis.

\section{A metallicity dependent reference function} 

\begin{figure*}
	\includegraphics[width=2\columnwidth]{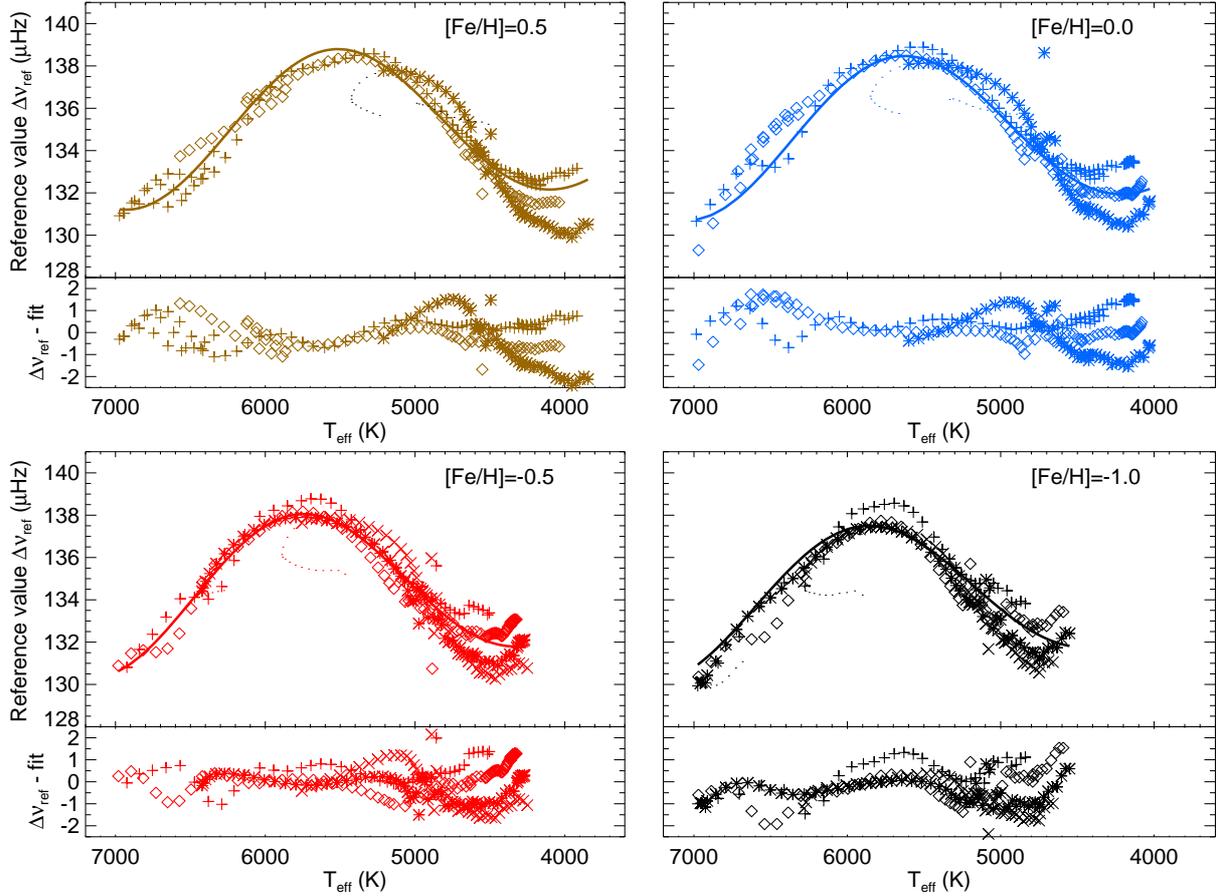}
    \caption{Each panel shows the reference values with the reference function (solid line) for one metallicity in the top part with residuals in the bottom part. The small dots indicate the models that
have not been included in the current analysis (see text for more details). The
residual value of the outlier in [Fe/H]~=~0.0 at $T_{\rm eff} \approx 4700$~K
is off the scale and not shown.}
    \label{fig:fit2}
\end{figure*}

From $\Delta\nu$ computed
as described above, and the known mass and radius of the model it is
possible to compute the value that $\Delta\nu_{\odot}$ should have for each
model to reproduce the correct mean density. We call this $\Delta\nu_{\rm ref}$.
These values, for our set of YREC models, are plotted in Fig.~\ref{fig:fit2} 
as a function
of $T_{\rm eff}$.  The reference values as a function of $T_{\rm eff}$ take the
shape of one half of a sinusoid with a decreasing amplitude. Hence, we fitted a
damped sinusoid using a Levenberg-Marquardt least-squares minimization. We then iteratively changed the coefficients of the damped
sinusoid to depend on [Fe/H]. This results in the following fit:
\begin{equation}
\Delta\nu_{\rm{ref}}=\rm{A} \cdot e^{\lambda \rm{T_{eff}}/10^4K} \cdot (\cos(\omega \cdot \rm{T_{eff}}/10^4K+\phi))+B	
\label{eq:corrfunc2}
\end{equation}
Parameters $A$, $\lambda$, $\omega$, $\phi$, and $B$ of
equation~\ref{eq:corrfunc2} are listed in Table~\ref{tab:pars2}. This function
(Eq.~\ref{eq:corrfunc2}) should be used instead of $\Delta\nu_{\odot}$ to take
$T_{\rm eff}$ and [Fe/H] effects into account in the scaling relations. The
residuals between the reference values for each model and
Eq.~\ref{eq:corrfunc2} are shown in the lower sub-panel of each of the four
panels in Fig.~\ref{fig:fit2}. These residuals still show a dependence on mass,
which will be addressed in a forthcoming paper.

The robustness of this fit was examined using a Monte-Carlo approach in which
we perturbed the model input values 10\,000 times assuming typical
uncertainties of 80~K in temperature, 1\% in \dnu\ and 0.1 dex in [Fe/H]
\citep{dav16}. A new fit was computed for each realisation. The maximum
deviation between fits, calculated by dividing the fits, is typically 0.44\% with a standard deviation of 0.16\%,
i.e. nearly an order of magnitude smaller than the effect we aim to correct
(see Fig.~\ref{fig:fit}). Finally, $\Delta\nu_{\rm ref}$ computed for the
models with the Krishna Swamy $T-\tau$ relation are in line with the values for
the other models (see Fig.~\ref{fig:fit3}).

\subsection{Parameter range of the validity of the reference function}
The models that are incorporated in this study are shown with thick lines in
Fig.~\ref{fig:HRD}.  Low-mass (M=0.8 and M=1.0) models on the beginning of the main-sequence do not
follow the general trend of the other models. These models create loops in
$\Delta\nu_{\rm ref}$--$T_{\rm eff}$ relation (shown as small dots in
Fig.~\ref{fig:fit2}) and we excluded them from our current analysis.
Models with $T_{\rm{eff}} > 7000$~K were also excluded as such high-temperature
stars will not have an outer convection zone deep enough to excite oscillations. These two cuts left the end of the main sequence in the sample, while removing the early main sequence phases. We note that most oscillating MS stars are observed in the later phases due to biases in the photometric observations of solar-like oscillations towards hotter and more evolved stars on the MS \citep{chap11}.
Our models go down to \numax\ = 6 $\mu$Hz which occurs at $T_{\rm eff}$ between 3800~K and
4500~K, depending on [Fe/H] and mass.

We note that there are a few models that have reference values that deviate from the general
trend (see Fig.~\ref{fig:fit} and \ref{fig:fit2}). We currently do not know the
reason for these deviations. One possibility is that this is related to the
luminosity bump.

\begin{table}
	\centering
	\caption{Parameters of the correction function.}
	\label{tab:pars2}
	\begin{tabular}{lc} % four columns, alignment for each
		\hline
		A & 0.64[Fe/H] + 1.78  $\mu Hz$ \\
		$\lambda$ & $-$0.55[Fe/H] + 1.23  \\
		$\omega$ & 22.21 rad/K \\
		$\phi$ & 0.48[Fe/H] + 0.12 \\
		B & 0.66[Fe/H] + 134.92 $\mu Hz$ \\
		\hline
	\end{tabular}
\end{table}

\subsection{Comparison of reference functions}
\begin{figure}
	\includegraphics[width=\columnwidth]{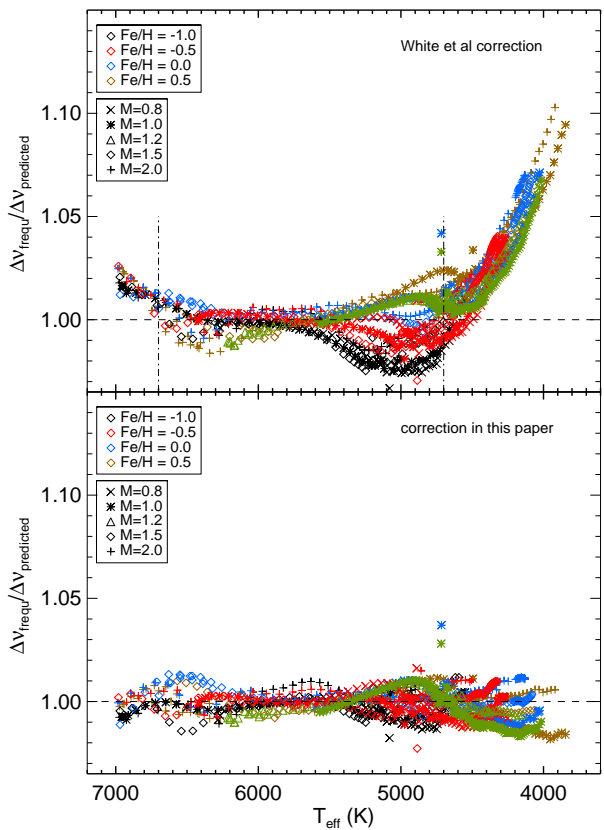}
    \caption{Same as Fig.~\ref{fig:fit} but now using the scaling relation (Eq.~\ref{dnu}) with corrections. The top panel uses the correction by~\citet{white11}. Vertical lines indicate the temperature range for which this correction function is valid. The bottom panel shows the result using the method proposed in this work where a reference function (Eq.~\ref{eq:corrfunc2}) is used instead of the solar reference value. The ratio for the models with the Krishna Swamy (KS) $T-\tau$ relation are indicated in green.}
    \label{fig:fit3}
\end{figure}

To justify the inclusion of a metallicity dependence in the reference function
we show that this 
a significant improvement in the derived stellar mean densities (and thus
masses and radii) using scaling relations. To do this, we compare the
performance of our $T_{\rm eff}$ and [Fe/H]-dependent reference function with
the quadratic function proposed by \citet{white11} that only depends on 
$T_{\rm eff}$.
We visualize the differences in the top panel of Fig.~\ref{fig:fit3}. The previously available correction works well in its defined range and for solar metallicity. However, for other metallicities systematic biases occur around 5000K. We mitigate these biases in this work, and at the same time we extend the range in which the reference works by about 1000K. This is about half of the previously available range.  
Additionally, we quantify the performance of the corrections by using a cross-validation technique: we fit our function
and the quadratic
temperature function of \citet{white11} 10 times, each time 10\% of
the models were randomly excluded from the procedure. We then predict the mean
density of the excluded models and compare the predicted values with the real
values using the coefficient of determination which is a goodness-of-fit measure defined as
\begin{equation}
R^2 \equiv 1-\frac{\sum_i(y_i-f_i)^2}{\sum_i(y_i-\bar{y})^2},
\end{equation}
where $y_i$ indicate the individual `observed' data points, $\bar{y}$ indicates the mean of the
observed data and $f_i$ indicates the predicted value \citep{weihs13}. This test is the standard method for model selection and evaluation in statistics. It evaluates how well our model is able to predict unseen values as opposed to describing values that have been fitted for. A value of $R^2=1$ means a perfect fit,
while a $R^2=0$ indicates that the fit does not reduce the variance. For the
reference function derived in this work we find a value of $R^2=0.88$. This value is
the same both for the whole parameter space investigated in this work, as well
as for the temperature range for which the quadratic function of \citet{white11}
was developed, i.e. 6700K $\ge \rm{T}_{\rm{eff}} \ge $4700K.
The $R^2$ value for the quadratic function recalibrated to our [Fe/H]=0 YREC models is
0.54 when applied to all metallicities. 

We note that the correction proposed by \citet{white11}  and the one
recalibrated to YREC models showed differences smaller than 0.5\%, indicating
that differences in stellar evolution codes (at least between YREC and ASTEC in
the relevant $T_{\rm eff}$ range) introduce a deviation of the same order as
the uncertainties in the measurements (see Section 3), which are significantly
smaller than the corrections to the $\Delta\nu$ scaling relation suggested by
\citet{white11} and in this work. 

\section{Conclusions}

We have developed a new reference function for use with the asteroseismic
scaling relation linking \dnu\ to the stellar mean density (Eq.~\ref{dnu}).
This reference function includes dependencies on $T_{\rm eff}$ and [Fe/H] and
increases the accuracy of masses and radii determined using the scaling
relations by a factor of 2.  It can immediately be used to estimate more accurate masses and
radii for the ten thousands of stars for which solar-like oscillations have
been observed by CoRot \citep{baglin06}, \textit{Kepler} \citep{bor08} and K2 \citep{haas14} and that are currently being
studied. Additionally, this can be applied to the many oscillators
expected to be observed by missions such as TESS \citep{ric15} and Plato
\citep{rau14}. While this functional form cannot fully capture the complex behavior of the \dnu\ deviations it has the advantage that it can be applied in a straightforward manner -- even without using grid-based modeling -- to extract masses and radii. Furthermore, it can also be used with existing grids that do not have oscillations computed.

This is the first reference function that takes metallicity (-1.0 dex $\le$
Fe/H $\le$ 0.5 dex) into account. This function is applicable to stars 
in different evolutionary states including (end of) main-sequence stars, subgiants and
cool red giant stars down to \numax\ = 6 $\mu$Hz, with the exception of low-mass main-sequence stars and
red-clump stars. Uncertainties in the data and differences in stellar evolution codes and model atmospheres impact the reference function at a level that is an order of magnitude smaller than the proposed improvement to the $\Delta\nu$ scaling relation. Hence, we consider the reference function proposed in this work to be widely applicable, i.e. to both models and observed data. A mass dependence still remains, especially on the red giant branch. This will be addressed in a forthcoming paper.

\section*{Acknowledgments}

The research leading to the presented results has received funding from the
European Research Council under the European Community's Seventh Framework
Programme (FP7/2007-2013) / ERC grant agreement no 338251 (StellarAges).
S.B. acknowledges partial support of NASA grant NNX13AE70G and NSF grant AST-1514676.

%%%%%%%%%%%%%%%%%%%%%%%%%%%%%%%%%%%%%%%%%%%%%%%%%%



\bibliographystyle{mnras}
\bibliography{scalingrelations} % if your bibtex file is called example.bib

%%%%%%%%%%%%%%%%%%%%%%%%%%%%%%%%%%%%%%%%%%%%%%%%%%

%%%%%%%%%%%%%%%%%%%%%%%%%%%%%%%%%%%%%%%%%%%%%%%%%%

% Don't change these lines
\bsp	% typesetting comment
\label{lastpage}
\end{document}